\documentclass[pre,showpacs]{revtex4}
\usepackage{epsf}

\begin{document}
\newcommand {\nl  } {\newline        } \newcommand {\nn  } {\nonumber     }
\newcommand {\noi } {\noindent       } \newcommand {\np  } {\newpage      }
\newcommand {\ov  } {\over        }    \newcommand {\pa  } {\partial      }
\newcommand {\e   } {\!+\!           } \newcommand {\m   } {\!-\!         }
\newcommand {\Bra } {\left\langle    } \newcommand {\Ket } {\right\rangle }
\newcommand {\lh  } {\left(          } \newcommand {\rh  } {\right)       }
\newcommand {\lv  } {\left[          } \newcommand {\rv  } {\right]       }
\newcommand {\lc  } {\left\{         } \newcommand {\rc  } {\right\}      }
\newcommand {\lp  } {\left.          } \newcommand {\rp  } {\right.       }
\newcommand {\de  } {\delta          } \newcommand {\om  } {\omega        }
\newcommand {\be  } {\beta           }
\newcommand {\rmm } {{\rm    m}      } \newcommand {\tin } {{\tilde{\rm n}}}
\newcommand {\ri  } {{\rm    i}      } \newcommand {\rn  } {{\rm    n}    }
\newcommand {\htom} {{\hat{\om }}}     \newcommand {\htpi} {{\hat{\pi }}}
\newcommand {\htrh} {{\hat{\rho}}}     \newcommand {\hty } {{\hat{y   }}}
\newcommand {\htm } {{\hat{m   }}}     \newcommand {\htq } {{\hat{q   }}}
\newcommand {\htx } {{\hat{x   }}}     
\newcommand {\htc } {{\hat{c   }}}     \newcommand {\htd } {{\hat{d   }}}
\newcommand {\vcc } {{\vec{c   }}}     \newcommand {\vn  } {{\vec{n   }}}
\newcommand {\vcr } {{\vec{r   }}}     \newcommand {\vs  } {{\vec{s   }}}
\newcommand {\vt  } {{\vec{t   }}}     \newcommand {\vz  } {{\vec{z   }}}
\newcommand {\cD  } {{\cal   D }}      \newcommand {\cF  } {{\cal   F }}       
\newcommand {\cG  } {{\cal   G }}      \newcommand {\cI  } {{\cal   I }}
\newcommand {\cM  } {{\cal   M }}      \newcommand {\cN  } {{\cal   N }}       
\newcommand {\cO  } {{\cal   O }}      \newcommand {\cQ  } {{\cal   Q }}       
\newcommand {\cS  } {{\cal   S }}      \newcommand {\cT  } {{\cal   T }}      
\newcommand {\cU  } {{\cal   U }}      \newcommand {\cW  } {{\cal   W }}      
\newcommand {\cX  } {{\cal   X }}      \newcommand {\cZ  } {{\cal   Z }}      
\newcommand {\bfA } {{\bf    A }}      \newcommand {\bfB } {{\bf    B }}
\newcommand {\bfC } {{\bf    C }}      \newcommand {\bfG } {{\bf    G }}
\newcommand {\BKK } {{\Bra\Ket_K }}    \newcommand {\BKm } {{\Bra a\Ket_m}}
\newcommand {\Tr  } {\mathop{\mbox{\rm Tr   }}}
\newcommand {\ext } {\mathop{\mbox{\rm ext  }}}
\newcommand {\ath } {\mathop{\mbox{\rm atanh}}}
\newcommand {\fns } {\footnotesize}
\newcommand {\hsc } {\hspace*{1cm} }
\newcommand {\ha  } {{1\over 2}}       \newcommand {\ev  } {\equiv        }
\newsavebox{\uuunit}
\sbox{\uuunit}
    {\setlength{\unitlength}{0.825em}
     \begin{picture}(0.6,0.7)
        \thinlines
        \put(0,0){\line(1,0){0.5}}
        \put(0.15,0){\line(0,1){0.7}}
        \put(0.35,0){\line(0,1){0.8}}
       \multiput(0.3,0.8)(-0.04,-0.02){12}{\rule{0.5pt}{0.5pt}}
     \end {picture}}
\newcommand {\Unity}{\mathord{\!\usebox{\uuunit}}}

\title{Critical Noise Levels for LDPC decoding}

\author{J. van~Mourik}
\author{D. Saad}
\affiliation{
The Neural Computing Research Group,  Aston University,
Birmingham B4 7ET, United Kingdom}

\author{Y. Kabashima}
\affiliation{
Department of Computational Intelligence and Systems Science,
Tokyo Institute of Technology, Yokohama 2268502, Japan}

\begin{abstract}
  We determine the critical noise level for decoding low density
  parity check error correcting codes based on the magnetization
  enumerator ($\cM$), rather than on the weight enumerator ($\cW$)
  employed in the information theory literature.  The interpretation
  of our method is appealingly simple, and the relation between the
  different decoding schemes such as typical pairs decoding, MAP, and
  finite temperature decoding (MPM) becomes clear.  In addition, our
  analysis provides an explanation for the difference in performance
  between MN and Gallager codes. Our results are more optimistic than
  those derived via the methods of information theory and are in
  excellent agreement with recent results from another statistical
  physics approach.
\end{abstract}
\pacs{89.70+c,89.90+n,05.50+q}

\maketitle

\section{Introduction}
%

The theory of error-correcting codes is based on the efficient introduction of
redundancy to given messages for protecting the information content against
corruption. The theoretical foundations of this area were laid by Shannon's
seminal work~\cite{shannon} and have been developing ever since. One of the main
results obtained in this field is the celebrated {\it channel coding theorem}
stating that there exists a code such that the average message error probability
$P_E$, when maximum likelihood decoding is used, can be made arbitrarily small
for sufficiently long messages below the {\it channel capacity}; and will
approach 1 above it. The channel coding theorem is based on unstructured random
codes and impractical decoders such as maximum likelihood \cite{viterbi} and
typical set decoding~\cite{Aji}. In the case of structured codes, the critical
code rate $R$ (message information content/length of the encoded transmission)
may lie below the channel capacity, commonly termed {\it Shannon's bound}, even
if optimal (and typically impractical) decoding methods are being used. The
proximity of the critical code rate to Shannon's limit provides an indication to
the theoretical limitations of a given code.

In 1963 Gallager \cite{gallager} proposed a coding scheme which involves sparse
linear transformations of binary messages that was forgotten soon after, in part
due to the success of convolutional codes~\cite{viterbi} and the computational
limitations of the time. Gallager codes have been recently rediscovered by
MacKay and Neal (MN), that independently proposed a  closely related
code~\cite{MacKay}. Variations of this family of codes, known as Low Density
Parity Check (LDPC) codes,  have displayed performance comparable (and sometimes
superior) to other state-of-the-art codes. This family of codes has been 
thoroughly investigated in the information theory (IT) literature (e.g.,
\cite{MacKay,Richardson,Aji}), providing a range of significant theoretical and
practical results.

In parallel to studies carried out in the IT community, a different approach
has been used to study LDPC codes, using the established methods of statistical
physics (SP). This analysis, relying mainly on the replica symmetric analysis of
diluted systems~\cite{MPV,nishimori_book}, offers an alternative to information
theory methods and has yielded some new results and insights
\cite{us_prl,cactus,montanari}. Due to the growing interest in LDPC codes and
their successful analysis via the methods of statistical physics, there is
growing interest in the relationship between IT and SP methods. As the two
communities investigate similar problems, one may expect that standard
techniques known in one framework would bring about new developments in the
other, and vice versa. Here we present a direct SP method to determine the
critical noise level of Gallager and MN error correcting codes, which allows us
to focus on the differences between the various decoding criteria and their use
for defining the critical noise level for which decoding is theoretically
feasible.

The paper is organized as follows: In section~\ref{sec:framework} we introduce
the general framework, notation and the quantities we focus on, while in
section~\ref{sec:SP} we will briefly describe the SP calculation.
Section~\ref{sec:qualitative} describes qualitatively the emerging picture of
the main quantities calculated for Gallager's code while the corresponding
picture for MN codes will be described in section~\ref{sec:MN}. Quantitative
results for the critical noise level will be presented in
section~\ref{sec:results} followed by conclusions.

\section{Regular Gallager and MN codes}
\label{sec:framework}

In a general scenario, the $N$ dimensional Boolean message $\vs^{o}\in\{0,1\}^N$
is encoded to the $M(\!>\!N)$ dimensional Boolean vector $\vt^o$, and
transmitted via a noisy channel, which is taken here to be a Binary Symmetric
Channel (BSC) characterized by an independent flip probability $p$ per bit;
other transmission channels may also be examined within a similar framework. 
At the other end of the channel, the corrupted codeword is decoded
utilizing the structured codeword redundancy.

The first type of error correcting code that we focus on here, is Gallager's
linear code~\cite{gallager}. Gallager's code is a low density parity check code
defined by the a binary $(M\m N)\!\times\!M$ matrix $\bfA=[\bfC_1|\bfC_2]$,
concatenating two very sparse matrices known to both sender and receiver, with
the $(M\m N)\!\times\!(M\m N)$ matrix $\bfC_2$ being invertible. The matrix
$\bfA$ has $K$ non-zero elements per row and $C$ per column, and the code rate
is given by $R\!=\!1\m C/K\!=\!1\m N/M$. Encoding refers to multiplying the
original message $\vs ^o$ with the $(M\!\times \!N)$ matrix ${\bf G}^T$ (where
${\bf G}\!=\! [\mbox{\boldmath{$\Unity$}}_N|{\bf C}_2^{\m 1}]$), yielding the
transmitted vector  $\vt ^o$. Note that all operations are carried out in (mod
2) arithmetic. Upon sending $\vec{t}^o$ through the binary symmetric channel
(BSC) with noise level $p$, the vector $\vcr=\vt ^o\e \vn ^o$ is received, where
$\vn ^o$ is the true noise.

Decoding is carried out by multiplying $\vcr $ by $\bfA$ to produce the syndrome
vector $\vz\!=\!\bfA \vcr $ ($=\bfA \vn ^o$, since $\bfA \bfG ^{T}={\bf 0}$).
In order to reconstruct the original message $\vs ^o$, one has to obtain an
estimate $\vn $ for the true noise $\vn ^o$. First we select all $\vn $ that
satisfy the parity checks $\bfA \vn =\bfA \vn ^o$:
\begin{equation}
\cI_{\rm pc}(\bfA ,\vn ^o)\equiv\{\vn ~|~\bfA \vn =\vz \},~~\mbox{and}~~
\cI^{\rm r}_{\rm pc}(\bfA ,\vn ^o)\equiv\{\vn \in \cI_{\rm pc}(\bfA,\vn ^o)~|~
\vn\neq \vn ^o\},
\label{Ipc_g}
\end{equation}
the (restricted) parity check set.

The second type of error correcting code that we focus on here is the
MacKay-Neal (MN) code~\cite{MacKay}. An MN code is a low density parity check
code defined by a binary $M\!\times\!(N\e M)$ matrix $\bfA=[\bfC_s|\bfC_n]$,
concatenating two very sparse matrices known to both sender and receiver, with
the $M\!\times\!M$ matrix $\bfC_n$ being invertible. The $M\!\times\!N$
matrix $\bfC_s$ has $K$ non-zero elements per row and $C$ per column, while
$\bfC_n$ has $L$ non-zero elements per row and column. The code rate is given
by $R\!=\!K/C\!=\!N/M$.
Encoding refers to multiplying the original message $\vs^o$ by the $(M\!\times
\!N)$ dense generator matrix ${\bf G}\!=\!\bfC_n^{\m1}\bfC_s$, yielding the
transmitted vector $\vt^o$. Note that all operations are carried out in
(mod2) arithmetic. Upon sending $\vt^o$ through the binary symmetric channel
(BSC) with noise level $p$, the vector $\vcr=\vt^o\e\vn^o$ is received, where
$\vn^o$ is the true noise.

Decoding is carried out by multiplying $\vcr$ by $\bfC_n$ to produce the
syndrome vector $\vz\!=\!\bfC_s\vs^o\e\bfC_n\vn^o\ev\bfA\vcc^o$, where $\vcc$
is the concatenated vector $(\vs,\vn)$.
In order to reconstruct the original message $\vs^o$, one has to obtain
estimates  $\vcc$ for the true signal and noise $\vcc^o$. First we select all
combinations of signal and noise $\vcc$ that satisfy the parity checks
$\bfA\vcc=\bfA\vcc^o$:
\begin{equation}
\cI_{\rm pc}(\bfA,\vcc^o)\ev\{\vcc~|~\bfA\vcc=\vz\},~~\mbox{and}~~
\cI^{\rm r}_{\rm pc}(\bfA,\vcc^o)\ev\{\vcc\in\cI_{\rm pc}(\bfA,\vcc^o)~|~
\vcc\neq\vcc^o\},
\label{Ipc_m}
\end{equation}
the (restricted) parity check set.

To unify notation for Gallager and MN codes, we will adopt the notation $\vcc^o$
for the original noise (and signal) vector, and $\vcc$ for the estimate of the
noise (and signal) vector.
Any general decoding scheme then consists of selecting a vector $\vcc^*$ from
$\cI_{\rm pc}({\bf A},\vcc^o)$, on the basis of some noise (and signal)
statistics criterion. Upon successful decoding $\vcc^o$ will be selected, while
a decoding error is declared when a vector $\vcc^*\in\cI^{\rm r}_{\rm
pc}(\bfA,\vcc^o)$ is selected. For each decoding scheme, the average {\em block
error probability}~\cite{iba}
\begin{equation}
P_e(p_s,p)=\Bra \Delta\left(~\mbox{a vector 
$\vcc\in \cI^{\rm r}_{\rm pc}(\bfA ,\vcc^o)$ is selected}~\right) 
~\Ket_{\bfA ,\vcc^o}\ 
\label{Pe}
\end{equation}
can be defined as a measure of error correcting ability for a given code
ensemble, where $\Delta(\cdot)$ is an indicator function returning $1$ if the
proposition of the argument is true and $0$, otherwise. 
For BSC, only the number of non-zero components characterizes the statistics 
of the noise. On the other hand, the signal bits in general have an equal
probability for being 0 and 1 (i.e. $p_s=\ha$), which implies that they have no
useful prior information for the estimation. In the following, we therefore
focus on decoding schemes based on the weight of a vector which is the average
sum of the noise components $w(\vcc)\ev{1\ov M}\sum_{j=1}^Mn_j$.
To obtain the error probability, one averages the indicator function over all
$\vcc^o$ vectors drawn from some distribution and the code ensemble $\bfA$ as
denoted by $\Bra.\Ket _{\bfA,\vcc^o}$.

Unfortunately, carrying out averages over the indicator function is difficult.
Therefore, the error probability~(\ref{Pe}) is usually upper-bounded by
averaging over the {\em number} of vectors $\vn$ obeying a certain condition on
the weight $w(\vn )$ which characterizes the employed decoding scheme.
Alternatively, one can find the average number of vectors with a given weight 
value $w$ from which one can construct a complete weight distribution of noise
vectors  $\vn$ in  $\cI^{\rm r}_{\rm pc}(\bfA,\vcc^o)$. From this distribution
one can, in principle, calculate a bound for $P_e$ and derive critical noise
values above which successful decoding cannot be carried out.

A natural and direct measure for the average number of states is the entropy 
of a system under the restrictions described above, that can be calculated 
via the methods of statistical physics.

It was previously shown (see e.g.~\cite{us_prl} for technical details) that this
problem can be cast into a statistical mechanics formulation, by replacing the
field $(\{0,1\},+{\rm mod(2)})$ by ($\{1,-1\},\times$), and by adapting the
parity checks correspondingly. The statistics of a noise vector $\vn $ is now
described by its magnetization $m(\vn )\equiv{1\over M}\sum_{j=1}^Mn_j$,
$(m(\vn)\in [1,-1])$, which is inversely linked to the vector weight in the
$[0,1]$ representation. Similarly, the statistics of a signal vector $\vs$ is
now described by its magnetization $m_s(\vs)\ev{1\ov M}\sum_{j=1}^Ms_j$,
$(m_s(\vs)\in [1,-1])$. With this in mind, we introduce the conditioned
magnetization enumerator, for a given  code and noise, measuring the noise
vector magnetization distribution in $\cI^{\rm r}_{\rm pc}({\bf A},\vn^o)$
\begin{equation}
\cM_{\bfA ,\vn ^o}(m)\equiv{1\over M}\ln\left[
\Tr_{\vn \in\cI ^{\rm r}_{\rm pc}(\bfA ,\vn ^o)}\delta(m(\vn )\m m)\right]~.
\label{MAn}
\end{equation}
To obtain the {\em magnetization enumerator} $\cM(m)$
\begin{equation}
\cM(m)=\Bra\frac{}{}~\cM_{\bfA,\vcc^o}(m)~\Ket_{\bfA,\vcc^o}~,
\label{M}
\end{equation} 
which is the entropy of the noise vectors in $\cI^{\rm r}_{\rm pc}(\bfA,\vn^0)$
with a given $m$, one carries out uniform explicit averages over all codes
$\bfA$ with given parameters $K,C$ (and $L$), and the weighted average over all
possible noise vectors generated by the BSC, (and all possible signal vectors)
i.e.,
\begin{eqnarray}
P(\vn^o)=\prod_j^M\lh(1\m p  )~\de(n^o_j\m1)+p  ~\de(n^o_j\e1)\rh,\\
P(\vs^o)=\prod_j^N\lh(1\m p_s)~\de(s^o_j\m1)+p_s~\de(s^o_j\e1)\rh,
\end{eqnarray}
with $p_s=\ha$.
It is important to note that, in calculating the entropy, the average quantity 
of interest is the magnetization enumerator rather than the actual number of
states. As physicists, this is the natural way to carry out the averages for
three main reasons: a) The entropy obtained in this way is believed to be {\em 
self-averaging}, i.e., its average value (over the disorder) coincides with its
{\em typical} value. b) This quantity is {\em extensive} and grows linearly with
the system size. c) This averaging distinguishes between {\em annealed}
variables that are averaged or summed for a given set of {\em quenched}
variables, that are averaged over later on. In this particular case, summation
over all $\vcc$ vectors is carried for a {\em fixed} choice of code $\bfA$ and
vector $\vcc^o$; averages over these variables are carried out at the next
level.

One should point out that in somewhat similar calculations, we showed that this
method of carrying out the averages provides more accurate results in comparison
to averaging over both sets of variables simultaneously~\cite{reliability}.

A positive magnetization enumerator, $\cM(m)\!>\!0$ indicates that there is an
exponential number of solutions (in $M$) with magnetization $m$, for typically
chosen $\bfA$ and $\vcc^o$,  while $\cM(m)\!\to\!0$ indicates that this number
vanishes as $M\!\to\!\infty$ (note that negative entropy is unphysical in
discrete systems).

Another important indicator for successful decoding is the overlap $\om$ between
the selected estimate $\vn^*$, and the true noise $\vn^o$:~$\om(\vn,\vn^o)
\equiv{1\ov M}\sum_{j=1}^Mn_jn^o_j$, $(\om(\vn,\vn^o)\in [-1,1])$, with $\om=1$
for successful (perfect) decoding. However, this quantity cannot be used for
decoding as $\vn^o$ is unknown to the receiver. The (code and noise dependent)
noise overlap enumerator is now defined as:
\begin{equation}
\cW _{\bfA,\vcc^o}(\om)\equiv{1\ov M}\ln\left[\Tr_{\vcc\in\cI^{\rm r}_{\rm pc}
(\bfA,\vcc^o)}\de(\om(\vn,\vn^o)\m\om)\right] \ ,
\end{equation}
and the average quantity being 
\begin{equation} 
\cW(\om)=\Bra\frac{}{}\cW_{\bfA,\vcc^o}(\om)\Ket_{\bfA,\vcc^o} \ .
\end{equation}
This measure is directly linked to the {\em weight enumerator} \cite{Aji},
although according to our notation, averages are carried out distinguishing
between annealed and quenched variables unlike the common definition in the IT
literature. However, as we will show below, the two types of averages provide
identical results {\em in this particular case}.

Similarly, for MN-codes one defines the signal magnetization and weight
enumerators as
\begin{eqnarray}
\cM_s(  m_s)&\ev&{1\ov N}\Bra\ln\left[\Tr_{\vcc\in\cI^{\rm r}_{\rm pc}
(\bfA,\vcc^o)}\de(  m(\vs      )\m  m_s)\right]\Ket_{\bfA,\vcc^o}\\
\cW_s(\om_s)&\ev&{1\ov N}\Bra\ln\left[\Tr_{\vcc\in\cI^{\rm r}_{\rm pc}
(\bfA,\vcc^o)}\de(\om(\vs,\vs^o)\m\om_s)\right]\Ket_{\bfA,\vcc^o}
\end{eqnarray} 
In what follows, we perform all calculations as if both $m$ and $\om$ (and $m_s$
and $\om_s$ for MN-codes), are constrained to particular values. As we will
show, omitting a constraint in the final expressions can then easily be done by
assigning the zero value to the corresponding Lagrange multiplier.

\section{The statistical physics approach}
\label{sec:SP}

Quantities of the type $\cQ(c)=\Bra\cQ_y(c)\Ket_y$, with
$\cQ_y(c)={1\ov M}\ln\left[{\cal Z}_y(c)\right]$ and
$\cZ_y(c)\equiv\Tr_x~\delta(c(x,y)\m Mc)$,  are very common in the SP of
disordered systems; the macroscopic order parameter $c(x,y)$ is fixed to a
specific value and may depend both on the disorder $y$ and on the microscopic
variables $x$. Although we will not prove this here, such a quantity is
generally believed to be {\em self-averaging} in the large system limit, i.e.,
obeying a probability distribution  $P\left(\cQ_y(c))=\delta(\cQ_y(c)-\cQ(c))
\right)$. The direct calculation of $\cQ(c)$ is known as a {\em quenched}
average over the disorder, but is typically hard to carry out and requires using
the replica method~\cite{nishimori_book}. The replica method makes use of the
identity $\Bra\ln\cZ\Ket=\Bra\ \lim_{\rn \to 0}[\cZ^\rn \m 1]/\rn \ \Ket$,
by calculating averages over a product of partition function replicas. Employing
assumptions about replica symmetries and analytically continuing the variable
$\rn $ to zero, one obtains solutions which enable one to determine the state of
the system.

To simplify the calculation, one often employs the so-called {\em annealed}
approximation, which consists of performing an average over $\cQ_y(c)$ first,
followed by the logarithm operation. This avoids the replica method and provides
(through the convexity of the logarithm function) an upper bound to the quenched
quantity:
\begin{equation}
Q_a(c)\equiv{1\ov M}\ln[\Bra \cZ_y(c)\Ket _y]~\geq~
Q_q(c)\equiv{1\ov M}\Bra \ln[\cZ_y(c)]\Ket _y=\lim_{\rn \!\to\!0}
{\Bra\cZ_y^\rn (c)\Ket_y\m 1\ov \rn M}~.
\label{Qaq}
\end{equation}

The technical details of the calculation are similar to those in~\cite{us_prl}. It turns
out that it is useful to perform the gauge transformation $c_j\!\to\!c_jc^o_j$,
such that the averages over the code $\bfA $ and noise/signal $\vcc^o$ can be
separated, $\cW_{\bfA,\vcc^o}$ becomes independent of $\vcc^o$, leading to an
equality between the quenched and annealed results, $\cW(m)=\cM_a(m)|_{p=0}=
\cM_q(m)|_{p=0}$. For any finite noise value $p$ one should multiply 
$\exp[\cW(\om)]$ by the probability that a state obeys all parity checks
$\exp[-K(\om,p)]$ given an overlap $\om$ and a noise level $p$~\cite{Aji}. In
calculating $\cW(\om)$ and $\cM_{a/q}(m)$, the $\de$-functions fixing $m$ and
$\om$, are enforced by introducing Lagrange multipliers $\htm$ and $\htom$.

Carrying out the averages explicitly one then employs the saddle point method to
extremize the averaged quantity with respect to the parameters introduced while 
carrying out the calculation. These lead, in both quenched and annealed
calculations, to a set of saddle point equations that are solved either
analytically or numerically to obtain the final expression for the averaged
quantity (entropy).

The final expressions for the annealed entropy per noise degree of freedom for
Gallager codes, under both overlap ($\om$) and magnetization ($m$) constraints,
are of the form:
\begin{equation}
\cQ_a=-{C\ov K}\!\left(\ln(2)\e(K\!\m\!1)\ln[1\e c_1^K]\right)\e\ln\!\Bra
\Tr_{n=\pm1}\exp(n(\htom\e\htm n^o))(1\e nc_1^{K\m1})^C\Ket_{n^o}
- (\htom\om \e \htm m)~,
\label{Qa_g}
\end{equation} 
where $c_1$ has to be obtained from the saddle point equation ${\partial\cQ_a
\ov\partial c_1}=0$. Similarly, the final expression in the quenched
calculation, employing the simplest replica symmetry
assumption~\cite{nishimori_book}, is of the form:
\begin{eqnarray}
\cQ _q\!&=&- C\!\!\int\!\!dxd\htx \ \pi(x)\htpi(\htx)~\ln[1\e x\htx]\e
{C\ov K}\int\!\left\{\prod_{k=1}^Kdx_k\pi(x_k)\right\}\ln\left[\ha
\left(1\e\!\prod_{k=1}^K\!x_k\right)\right]\nonumber\\
&&+\!\int\!\left\{\prod_{c=1}^Cd\htx_c\htpi 
(\htx_c)\right\}\Bra\!\ln\left[\Tr_{n=\pm1}\exp(n(\hat{\om}\e\hat{m}n^o))
\prod_{c=1}^C(1\e n\htx_c) 
\right]\!\Ket_{n^o} - (\htom\om \e \htm m)~. 
\label{Qq_g}
\end{eqnarray}
The probability distributions $\pi(x)$ and $\htpi(\htx)$ emerge from the
calculation; the former represents a probability distribution with respect to
the noise vector local magnetization~\cite{TAP_book}, while the latter relates
to a field of conjugate variables which emerge from the introduction of
$\de$-functions while carrying out the averages (for details see~\cite{us_prl}).
Their explicit forms are obtained from the functional saddle point equations
${\de\cQ_q\over\de \pi(x)}$, ${\de{\cQ}_q\ov\de\htpi(\htx)}=0$, and all
integrals are from $\m1$ to 1.

The final expressions for the annealed entropy per noise degree of freedom for
MN-codes, under both signal and noise overlap ($\om,\om_s$) and magnetization
($m,m_s$) constraints, are of the form:
\begin{eqnarray}
\cQ_a&=&-\lh \log(2)\e (K\!+\!L\!-\!1)\ln[1+c_1^Kd_1^L]\rh
        -R(\htm_sm_s\e\htom_s\om_s)-(\htm m\e\htom\om)                     \nn\\
  &&+R\ln\Bra\Tr_{s=\pm1}\exp\lh s(\htom_s+\htm_s~s^o)\rh(1+s\htc_1)^C\Ket_{s^o}
    + \ln\Bra\Tr_{n=\pm1}\exp\lh n(\htom  +\htm  ~n^o)\rh(1+n\htd_1)^L\Ket_{n^o}
\label{Qa_m}
\end{eqnarray}
where $c_1,~d_1$ have to be obtained from the saddle point equations
${\pa\cQ_a\ov\pa c_1},{\pa\cQ_a\ov\pa d_1}=0$.
Similarly, the final expression in the quenched calculation, employing the
simplest replica symmetry assumption~\cite{nishimori_book}, is of the form:
\begin{eqnarray}
\cQ_q&=&  \int\prod_{k=1}^Kdx_k~\pi(x_k)\prod_{l=1}^Ldy_l~\rho(y_l)~
              \ln\lv\ha\lh1\e \prod_{k=1}^Kx_k\prod_{l=1}^Ly_l\rh\rv
        -R(\htm_sm_s\e\htom_s\om_s)
        - (\htm  m  \e\htom  \om  )                                        \nn\\
     & &-K\int dxd\htx~\pi (x)\htpi(\htx)~\ln[1\e x\htx]
        +R\int\prod_{c=1}^Cd\htx_c~\htpi(\htx_c)\Bra\ln\lv\Tr_{s=\pm1}
          \exp(s(\htom_s+\htm_ss^o))\prod_{c=1}^C(1\e s\htx_c)\rv\Ket_{s^o}\nn\\
     & &-L\int dyd\hty~~\rho(y)\htrh(\hty)~~\ln[1\e y\hty]
        +~~\int\prod_{l=1}^Ld\hty_l~\htrh(\hty_l)~~\Bra\ln\lv\Tr_{n=\pm1}
          \exp(n(\htom ~+\htm ~n^o )\prod_{l=1}^L(1\e n\hty_l~)\rv\Ket_{n^o}
\label{Qq_m}
\end{eqnarray}
The probability distributions $\pi(x),\rho(y)$ and $\htpi(\htx),\htrh(\hty)$
emerge from the calculation; the former represent probability distributions
with respect to the signal/noise vector local magnetizations~\cite{TAP_book},
while the latter relate to fields of conjugate variables which emerge from the
introduction of $\de$-functions while carrying out the averages (for details
see~\cite{us_prl}).
Their explicit forms are obtained from the functional saddle point equations
${\de\cQ_q\over\de \pi(x)},{\de{\cQ}_q\ov\de\htpi(\htx)},
 {\de\cQ_q\over\de\rho(y)},{\de{\cQ}_q\ov\de\htrh(\hty)}=0$, and all integrals
are from $\m1$ to 1.

Enforcing a $\de $-function corresponds to taking $\htom,\htm,\htom_s,\htm_s$
such that ${\pa\cQ_{a/q}\ov\pa\htom  },{\pa\cQ_{a/q}\ov\pa\htm  }, 
           {\pa\cQ_{a/q}\ov\pa\htom_s},{\pa\cQ_{a/q}\ov\pa\htm_s}=0$,
while not enforcing it corresponds to putting $\htom,\htm,\htom_s,\htm_s$ to 0.
Since $\om,m,\om_s,m_s$, follow from ${\pa\cQ_{a/q}\ov\pa\htom},
{\pa\cQ_{a/q}\ov\pa\htm},{\pa\cQ_{a/q}\ov\pa\htom_s},{\pa\cQ_{a/q}\ov\pa\htm_s}
\!=\!0$, all the relevant quantities can be recovered with appropriate choices
of $\htom,\htm,\htom_s,\htm_s$.
%
\begin{figure}[h]
\setlength{\unitlength}{0.88mm}
\begin{picture}(140,100)
\put(  0, 50){\epsfysize=50\unitlength\epsfbox{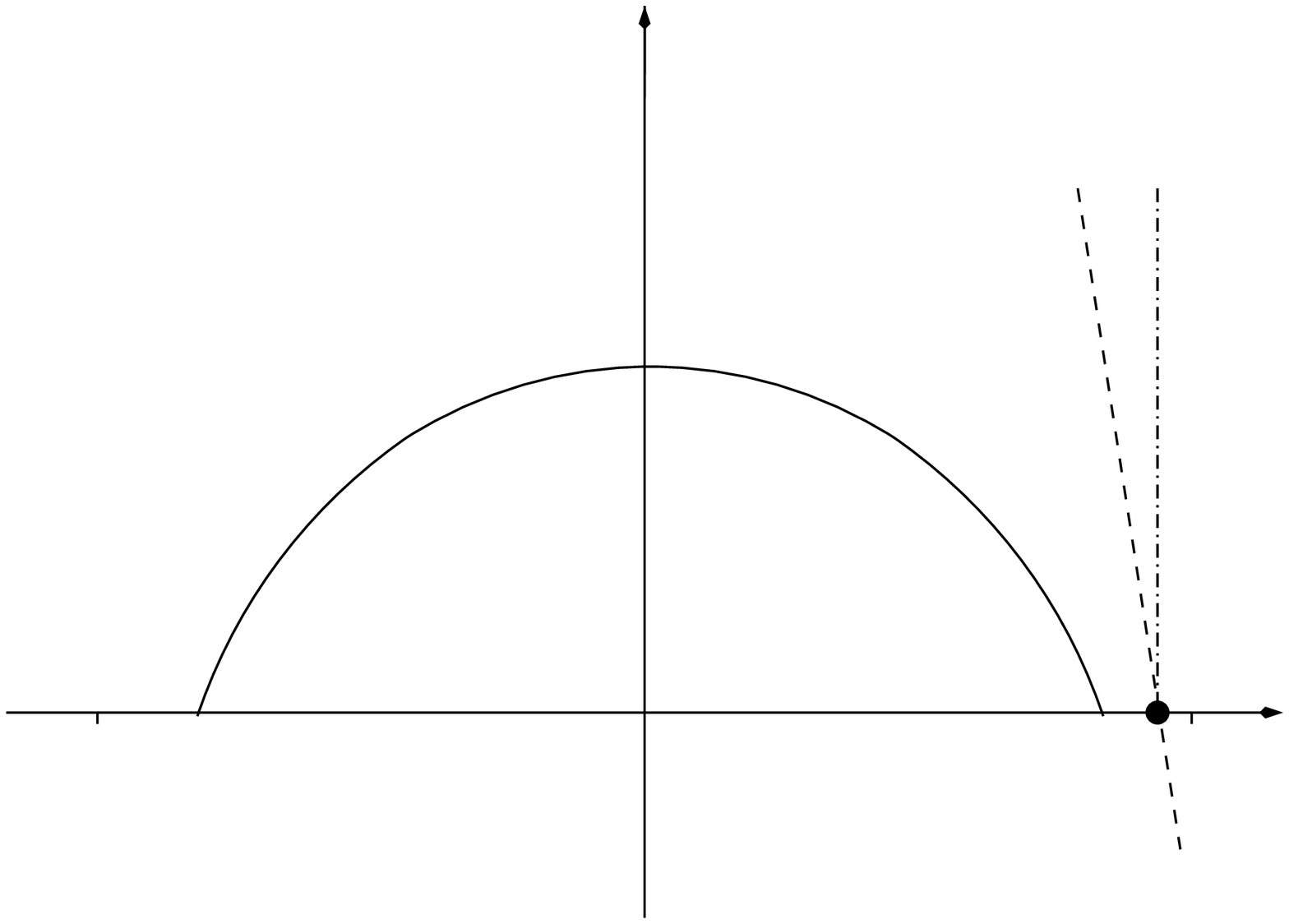}}
\put(  5, 90){\mbox{\fns \boldmath ${\rm a)}~p\!<\!p_c$}}
\put( 35, 95){\mbox{\fns \boldmath $\cM(m)$}}
\put( 40, 58){\mbox{\fns \boldmath $m$}}
\put( 50, 58){\mbox{\fns \boldmath $m_{\e}(p)$}}
\put(  2, 58){\mbox{\fns \boldmath $-1$}}
\put( 64, 58){\mbox{\fns \boldmath $1$}}
\put( 70, 50){\epsfysize=50\unitlength\epsfbox{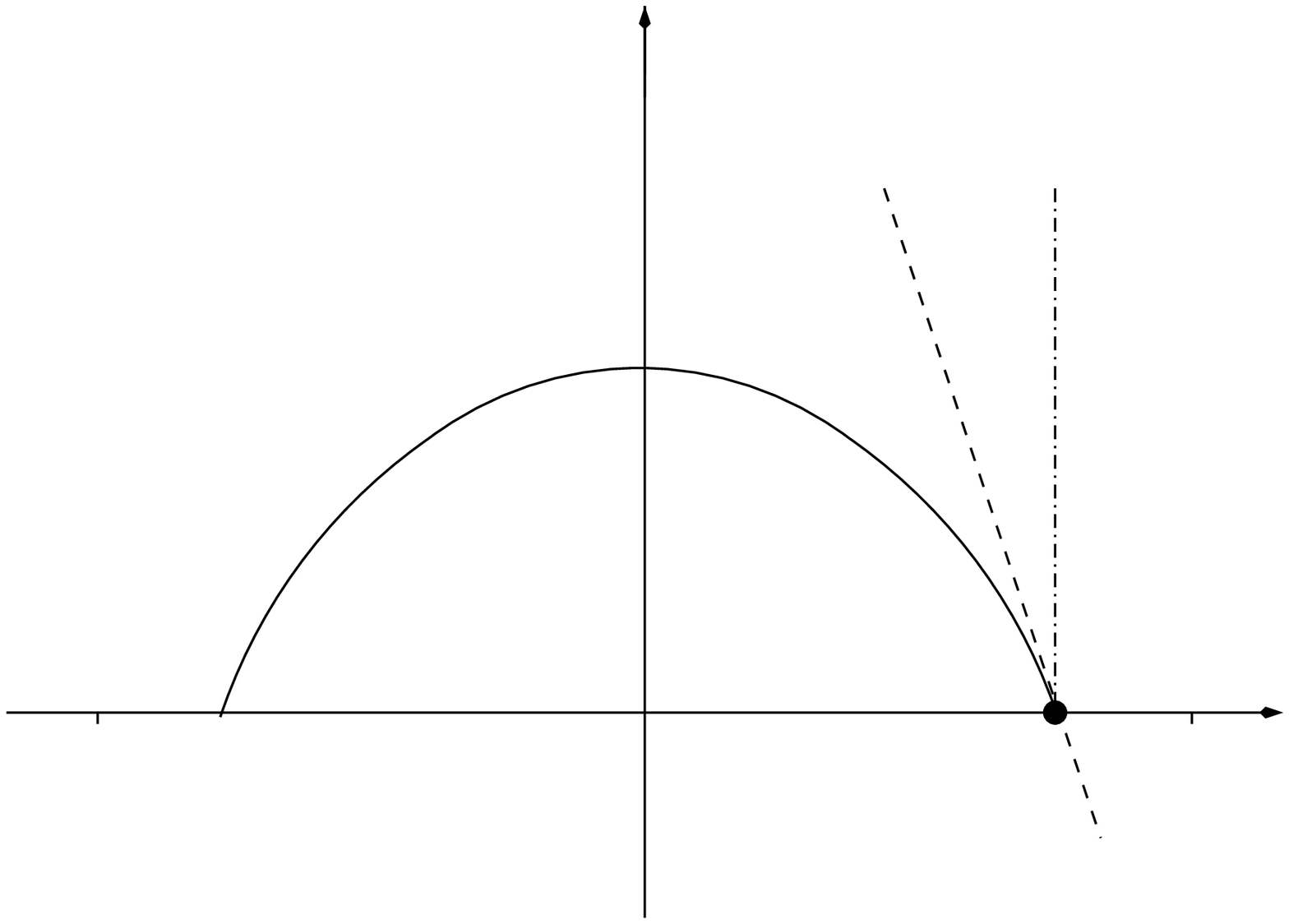}}
\put( 75, 90){\mbox{\fns \boldmath ${\rm b)}~p\!=\!p_c$}}
\put(105, 95){\mbox{\fns \boldmath $\cM(m)$}}
\put(110, 58){\mbox{\fns \boldmath $m$}}
\put(120, 58){\mbox{\fns \boldmath $m_{\e}(p)$}}
\put( 72, 58){\mbox{\fns \boldmath $-1$}}
\put(134, 58){\mbox{\fns \boldmath $1$}}
\put( 35,  0){\epsfysize=50\unitlength\epsfbox{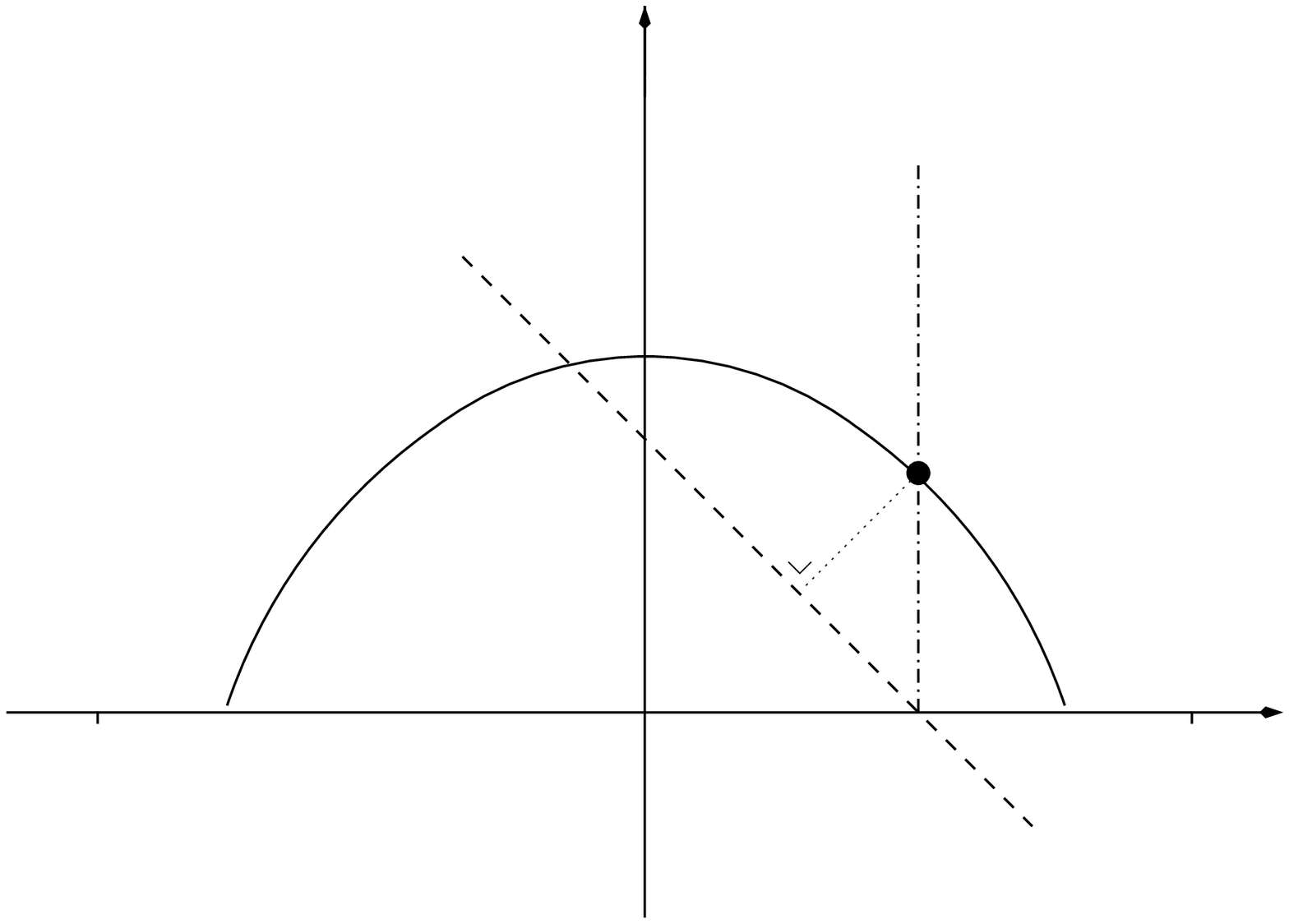}}
\put( 40, 40){\mbox{\fns \boldmath ${\rm c)}~p\!>\!p_c$}}
\put( 70, 45){\mbox{\fns \boldmath $\cM(m)$}}
\put( 75,  8){\mbox{\fns \boldmath $m$}}
\put( 85,  8){\mbox{\fns \boldmath $m_{\e}(p)$}}
\put( 37,  8){\mbox{\fns \boldmath $-1$}}
\put( 99,  8){\mbox{\fns \boldmath $1$}}
\put( 63,80){\mbox{\fns \boldmath $m_0(p)$}}
\put(127,80){\mbox{\fns \boldmath $m_0(p)$}}
\put( 85,30){\mbox{\fns \boldmath $m_0(p)$}}
\end{picture}
\caption{
  The qualitative picture of $\cM(m)\!\geq\!0$ (solid lines) for
  different values of $p$. For MAP, MPM and typical set decoding, only
  the relative values of $m_{\e}(p)$ and $m_{0}(p)$ determine the
  critical noise level. Dashed lines correspond to the energy contribution of
  $-\beta F$ at Nishimori's condition ($\beta=1$). The states with the lowest
  free energy are indicated by a point $\bullet$.
  {\bf a)} Sub-critical noise levels $p\!<\!p_{c}$, where $m_{\e}(p)\!<\!m_{0}
  (p)$, there are no solutions with higher magnetization than $m_{0}(p)$, and
  the correct solution has the lowest free energy.
  {\bf b)} Critical noise level $p\!=\!p_c$, where $m_{\e}(p)\!=\!m_{0}(p)$. The
  minimum of the free energy of the sub-optimal solutions is equal to that of
  the correct solution at Nishimori's condition.
  {\bf c)} Over-critical noise levels $p\!>\!p_c$ where many solutions have a
  higher magnetization than the true typical one. The minimum of the free
  energy of the sub-optimal solutions is lower than that of the
  correct solution.
}\label{fig:1}
\end{figure}
%
\section{Qualitative picture}
\label{sec:qualitative}
We now discuss the qualitative behaviour of $\cM(m)$, and the interpretation of
the various decoding schemes. To obtain separate results for  $\cM(m)$ and
$\cW(m)$ we calculate the results of Eqs.(\ref{Qa_g}) and (\ref{Qq_g}) (and Eqs.
(\ref{Qa_m}) and (\ref{Qq_m})), corresponding to the annealed and quenched
cases respectively, setting $\htom=0$ to obtain $\cM(m)$ and $\htm\!=\!0$ to
obtain $\cW(\om)$ (that becomes $\cM(m)|_{p=0}$ after gauging).
In Fig. \ref{fig:1}, we have qualitatively plotted the resulting function
$\cM(m)$ for relevant values of $p$. $\cM(m)$ (solid line) only takes positive
values in the interval $[m_{\m}(p),m_{\e}(p)]$; for even $K$, $\cM(m)$ is an
even function of $m$ and $m_{\m}(p)\!=\!-m_{\e}(p)$. The maximum value of
$\cM(m)$ is always $(1\m R)\ln(2)$ for Gallager codes, and $R\ln(2)$ for MN
codes. The true noise $\vn^o$ has (with probability 1) the typical magnetization
of the BSC: $m(\vn^o)\!=\!m_0(p)\!=\!1\m2p$ (dashed-dotted line).

The various decoding schemes can be summarized as follows:
\begin{itemize}
\item{\bf Maximum likelihood (MAP) decoding} -
  minimizes the block error probability~\cite{iba} and consists of
  selecting the $\vn $ from $\cI_{\rm pc}(\bfA ,\vn ^0)$ with the highest
  magnetization. Since the probability of error below $m_{\e}(p)$ vanishes,
  $P(\exists\vn\in\cI^{\rm r}_{\rm pc}: m(\vn )\!>\! m_{\e}(p))\!=\!0$, and
  since $P(m(\vn^o)\!=\!m_0(p))\!=\!1$, the critical noise level $p_c$ is
  determined by the condition $m_{\e}(p_c)\!=\!m_0(p_c)$. The selection process
  is explained in Fig.\ref{fig:1}(a)-(c).
\item{\bf Typical pairs decoding} -
  is based on randomly selecting a $\vn $ from $\cI_{\rm pc}$ with $m(\vn)=m_0
  (p)$~\cite{Aji}; an error is declared when $\vn ^0$ is not the only element of
  $\cI_{\rm pc}$. For the same reason as above, the critical noise level $p_c$
  is determined by the condition $m_{\e}(p_c)\!=\!m_0(p_c)$.  
\item{\bf Finite temperature (MPM) decoding} - An energy $-Fm(\vn )$
  (with $F\!=\!\ha\ln({1-p\ov p})$) according to Nishimori's condition
  (corresponding to the selection of an accurate prior within the
  Bayesian framework).  is attributed to each $\vn \in\cI_{\rm pc}$,
  and a solution is chosen from those with the magnetization that
  minimizes the free energy~\cite{us_prl}.  This procedure is known to
  minimize the {\em bit error probability} \cite{iba}.  Using the
  thermodynamic relation $\cF=\cU-{1\ov\beta}\cS$, $\beta$ being the
  inverse temperature (Nishimori's condition corresponds to setting
  $\beta\!=\!1$), the free energy of the sub-optimal solutions is
  given by $\cF(m)\!=\!-Fm\m{1\ov\beta}\cM(m)$ (for
  $\cM(m)\!\geq\!0$), while that of the correct solution is given by
  $-Fm_0(p)$ (its entropy being 0). The selection process is explained
  graphically in Fig.\ref{fig:1}(a)-(c).  The free energy differences
  between sub-optimal solutions relative to that of the correct
  solution in the current plots, are given by the orthogonal distance
  between $\cM(m)$ and the line with slope $-\beta F$ through the
  point $(m_0(p),0)$. Solutions with a magnetization $m$ for which
  $\cM(m)$ lies above this line, have a lower free energy, while those
  for which $\cM(m)$ lies below, have a higher free energy. Since
  negative entropy values are unphysical in discrete systems, only
  sub-optimal solutions with $\cM(m)\!\geq\!0$ are considered. The
  lowest $p$ value for which there are sub-optimal solutions with a
  free energy equal to $-Fm_0(p)$ is the critical noise level $p_{c}$
  for MPM decoding. In fact, using the convexity of $\cM(m)$ and
  Nishimori's condition, one can show that the slope $\pa\cM(m)/\pa
  m\!>\!-\beta F$ for any value $m\!<\!m_o(p)$ and any $p$, and equals
  $-\beta F$ only at $m\!=\!m_o(p)$; therefore, the critical noise
  level for MPM decoding $p\!=\!p_c$ is identical to that of MAP, in
  agreement with results obtained in the information theory
  community~\cite{MacKay_thrm}.
  
  The statistical physics interpretation of finite temperature decoding
  corresponds to making the specific choice for the Lagrange multiplier
  $\htm\!=\!\beta F$ and considering the free energy instead of the entropy.
  In earlier work on MPM decoding in the SP framework~\cite{us_prl}, negative
  entropy values were treated by adopting different replica symmetry
  assumptions, which effectively result in changing the inverse temperature,
  i.e., the Lagrange multiplier $\htm$. This effectively sets $m\!=\!m_{\e}(p)$,
  i.e. to the highest value with non-negative entropy. The sub-optimal states
  with the lowest free energy are then those with $m\!=\!m_{\e}(p)$.
\end{itemize}
The central point in all decoding schemes, is to select the correct solution
only on the basis of its magnetization. As long as there are no sub-optimal
solutions with the same magnetization, this is in principle possible. As shown
here, all three decoding schemes discussed above, manage to do so. To find
whether at a given $p$ there exists a gap between the magnetization of the 
correct solution and that of the nearest sub-optimal solution, just requires
plotting $\cM(m)(>\!0)$ and $m_0(p)$, thus allowing a graphical determination of
$p_c$. Since MPM decoding is done at Nishimori's temperature, the simplest
replica symmetry assumption is sufficient to describe the thermodynamically
dominant state~\cite{nishimori_book}. At $p_c$ the states with $m_{\e}(p_c)\!=
\!m_0(p_c)$ are thermodynamically dominant, and the $p_c$ values that we obtain
under this assumption are exact.
%
\section{MN codes - an alternative view}
\label{sec:MN}
For MN codes there is a way to obtain the {\em exact} expression for $\cM$, in
the case of unbiased messages, by employing a single highly plausible
assumption. We first note that every the parity check bit
$z_{<>}=s^o_{i_1}..s^o_{i_K}n^o_{j_1}..n^o_{j_L}$ is made up of a combination of
$K$ unbiased (i.e. $p_s=\ha$) signal bits, and $L$ biased (i.e. $p\neq\ha$)
noise bits. As a result, every syndrome element $z_{<>}$ is unbiased
independently of the noise bit statistics. It is therefore plausible to assume
that the noise bit statistics (i.e. $p$) have no influence on the distribution
of the parity check bits $z_{<>}$, and therefore on $\cM$ (which only depends on
the true noise through the $z_{<>}$). If this assumption is satisfied, one can
invoke Nishimori's condition to obtain an exact expression for $\cM$.
\nl
Independently of the assumption, Nishimori's condition gives the following
identity for the thermodynamically dominant state:
\begin{equation}
\lp{\pa\cM(m)\ov\pa m}\right|_{m=m_o(p)}=-F(p)=-\ha\ln\left({1\m p\ov p}\right)=
-\ha\ln\left({1\e m_o\ov 1\m m_o}\right)~.
\label{Nishi}
\end{equation}
Since states characterized by any magnetization value $m<m_0(p_t)$ will become
dominant for an appropriately chosen value of $p$, and since we assume that
$\cM$ is independent of $p$, the identity
\begin{equation}
{\pa\cM(m)\ov\pa m}=-\ha\ln\lh{1\e m\ov 1\m m}\rh~,
\end{equation}
must hold for any value of $m$. Furthermore, the maximum of $\cM(m)$ is reached
at $m=0$ with $\cM(0)=R\ln(2)$, and we have that
\begin{equation}
\cM(m)=\cM(0)\m\ha\int_0^m du\ln\lh{1\e u\ov1\m u}\rh=\ln(2)
\lv R-1+H_2\lh{1+m\ov2}\rh\rv~,
\label{MNS}
\end{equation}
where $H_2(p)$ is the binary entropy per bit for vectors with bias $p$. Hence,
under this assumption, we do not only obtain the exact expression for $\cM(m)$,
but we see that the critical noise level $p_c$ is given by $R=1-H_2(p_c)$,
saturating Shannon's bound for this type of codes!

Unfortunately, the assumption can not be verified easily without the replica
method. To verify whether indeed ${\pa\cM(m)\ov\pa p}=0$, we have to take
the derivative of expression~(\ref{Qq_m}) (setting $\htom=\htom_s=\htm_s=0$)
with respect to $p$. It turns out that $\cM$ is only independent of $p$, when
$\rho(\hty)$ is an even function of $\hty$, which in turn requires that
$\rho(y)$ and $\pi(x)$ are even functions of their arguments. Numerical analysis
shows, that this is the case for any $K\geq3$ or $K=2,~L\geq3$, while not so for
$K=1$ or $K=L=2$. This result is consistent with those reported in
\cite{us_prl}, i.e. that typical MN codes with $K\geq3$ or $K=2,~L\geq3$ do
saturate Shannon's bound, while those with $K=1$ and $K=L=2$ do not.

Intuitively this result can be understood in the following way. There are $M$
parity check bits and only $N(<M)$ signal bits, such that parity check bits,
although individually unbiased, are not uncorrelated. These correlations do seem
to have an effect on $\cM(m)$ for $K=1$ and $K=L=2$, while for $K\geq3$ and
$K=2,~L\geq3$ the signal bits seem to be ``scrambled'' enough in the parity
checks for the correlations to be insignificant. Note that this argument does
not hold for Gallager codes and MN codes with biased messages, where the parity
check bits exclusively comprise biased bits, and are therefore biased
themselves. They only become unbiased as $K\to\infty$ for Gallager codes (for
which it was already reported in the literature \cite{MacKay} that such codes
can saturate Shannon's bound), and for $K\to\infty$ or $L\to\infty$ for MN
codes.

In fact, numerical analysis reveals that for $K\geq3$ and for $K=2,~L\geq3$
we have that $\rho(\hty)=\de(\hty)$, $\rho(y)=\de(y)$, $\pi(x)=\de(x)$ at least
up to $m_+(p)=m_0(p_t)$ which is independent of $p$. This allows us to calculate
$\cM$ analytically from expression (\ref{Qq_m}), and we recover, as expected,
the exact expression (\ref{MNS}).

For $K=1$ or $K=L=2$, like in the case of Gallager codes, one can only obtain
$m_+(p)$ numerically. The results of this procedure are presented in the next
section. Furthermore, for $K=1$ and for $K=L=2$, we find that spontaneously
$m_s\neq0$ for some values of $p<p_c$, when no restriction is enforced (i.e. for
$\htm_s=0$). This implies that one may improve the decoding performance by
imposing the condition of unbiased signal (similar to the conditions for typical
set decoding), i.e. by adjusting the Lagrange multiplier $\htm_s$ such that
$m_s=0$. Unfortunately, this only happens for values of $p$ for which there is
an exponential number of sub-optimal solutions $\vcc\in\cI^{\rm r}_{\rm
pc}(\bfA,\vcc^o)$ with the same weight as $\vcc^o$, and imposing this constraint
on the signal estimator only reduces this number, leaving it nevertheless,
exponential.

It was shown~\cite{book} that MN codes in principle contain sufficient
information to saturate Shannon's bound for unbiased messages. For codes with
$K=1$, or $K=L=2$, some of this information is wasted in a region where
errorless decoding is impossible anyway, such that Shannon's bound is not
saturated. For codes with $K\geq3$, or $K=2,~L\geq3$, our analysis indicates
that all information is used optimally, and that Shannon's bound can be
theoretically saturated. Our argument also explains the relative importance of
the parameters $K$ and $L$ for the behaviour of the code in comparison with $C$.

\section{Critical noise level - results}
\label{sec:results}

\begin{figure}[h]
\setlength{\unitlength}{0.6mm}
\begin{picture}(140,60)
\put(- 9,  0){\epsfysize=60\unitlength\epsfbox{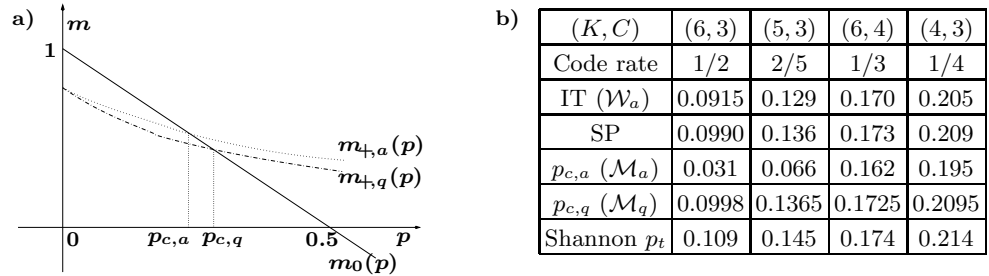}}
\put(-10, 55){\mbox{\fns \boldmath ${\rm a)}$}}
\put(- 3, 48){\mbox{\fns \boldmath $1$}}
\put(  2, 54){\mbox{\fns \boldmath $m$}}
\put(  2,  6){\mbox{\fns \boldmath $0$}}
\put( 75,  7){\mbox{\fns \boldmath $p$}}
\put( 20,  7){\mbox{\fns \boldmath $p_{c,a}$}}
\put( 32,  7){\mbox{\fns \boldmath $p_{c,q}$}}
\put( 60,  1){\mbox{\fns \boldmath $m_0(p)$}}
\put( 62, 27){\mbox{\fns \boldmath $m_{\e,a}(p)$}}
\put( 62, 20){\mbox{\fns \boldmath $m_{\e,q}(p)$}}
\put( 55,  6){\mbox{\fns \boldmath $0.5$}}
\end{picture}
\put(-43,55){\mbox{\fns \boldmath ${\rm b)}$}}
\put(-35,30){
\begin{tabular}{|c|c|c|c|c|}
\hline
$(K,C)$             & $(6,3)$ & $(5,3)$ & $(6,4)$ & $(4,3)$ \\
\hline
Code rate           & $1/2$   & $2/5 $  & $1/3 $  & $1/4 $  \\
\hline
IT ($\cW_a$)        & 0.0915  & 0.129   & 0.170   & 0.205   \\
\hline
SP                  & 0.0990  & 0.136   & 0.173   & 0.209   \\
\hline
$p_{c,a}$ ($\cM_a)$ & 0.031   & 0.066   & 0.162   & 0.195   \\
\hline
$p_{c,q}$ ($\cM_q)$ & 0.0998  & 0.1365  & 0.1725  & 0.2095  \\
\hline
Shannon $p_t$       & 0.109   & 0.145   & 0.174   & 0.214   \\
\hline
\end{tabular}
}
\caption{
 {\bf a)} Determining the critical noise levels $p_{c,a/q}$
  based on the function $\cM_{a/q}$ for Gallager codes and for MN codes with
  $K=1$ or $K=L=2$, a qualitative picture.
 {\bf b)} Comparison of different critical noise level ($p_c$) estimates for
  Gallager codes. Typical set decoding estimates have been obtained via the
  methods of IT~\cite{Aji}, based on having a unique solution to
  $\cW(m)\!=\!K(m,p_c)$, as well as using the methods of SP~\cite{KNM}. The
  numerical precision is up to the last digit for the current method. Shannon's
  limit denotes the highest theoretically achievable critical noise level $p_t$
  for any code~\cite{shannon}.
}
\label{fig:2}
\end{figure}

Some general comments can be made about the critical MAP (or typical set) 
values obtained via the annealed and quenched calculations. Since
$\cM_q(m)\leq\cM_a(m)$ (for given values of  $K$, $C$ ($L$) and $p$), we can
derive the general inequality $p_{c,q}\!\geq\!p_{c,a}$. For all $K$, $C$ ($L$)
values that we have numerically analyzed, for both annealed and quenched cases,
$m_{\e}(p)$ is a non increasing function of $p$, and $p_c$ is unique. The
estimates of the critical noise levels $p_{c,a/q}$, based on $\cM_{a/q}$, are
obtained by numerically calculating $m_{c,a/q}(p)$, and by determining their
intersection with $m_0(p)$. This is explained graphically in Fig.\ref{fig:2}(a).
As the results for MPM decoding have already been presented elsewhere
\cite{cactus}, we will now concentrate on the critical results $p_c$ obtained
for typical set and MAP decoding for Gallager codes; these are presented in
Fig.\ref{fig:2}(b), showing the values of $p_{c,a/q}$ for various choices of $K$
and $C$ compared with those reported in the literature.

>From the table it is clear that the annealed approximation gives a much more
pessimistic estimate for $p_c$. This is due to the fact that it overestimates
$\cM$ in the following way. $\cM_a(m)$ describes the combined entropy of $\vn$
and $\vn^o$ as if $\vn^o$ were thermal variables as well. Therefore,
exponentially rare events for $\vn ^o$ (i.e. $m(\vn ^o)\!\neq\!m_0(p)$) still
may carry positive entropy due to the addition of a positive entropy term from 
$\vn$. In a separate study~\cite{KNM} these effects have been taken care of by 
the introduction of an extra exponent; this is not necessary in the current
formalism as the quenched calculation automatically suppresses such
contributions. The similarity between the results reported here and those
obtained in~\cite{reliability} is not surprising as the equations obtained in
quenched calculations are similar to those obtained by averaging the upper-bound
to the reliability exponent using a methods presented originally by
Gallager~\cite{gallager}. Numerical differences between the two sets of results
are probably due to the higher numerical precision here.

We have also obtained the critical noise levels for some parameter choices
in MN codes. We only present the quenched (exact) values, and compare them
only with the highest theoretically achievable critical noise level $p_t$ for
any code~\cite{shannon}, as we are not aware of values obtained with other
methods in the literature. Note that although still strictly below $p_t$,
the critical noise levels $p_c$ for $K=L=2$ with increasing values of $C$
rapidly approach $p_t$ to within the current numerical precision.

\begin{figure}[h]
\setlength{\unitlength}{0.6mm}
\begin{picture}(140,60)
\put(- 9,  0){\epsfysize=60\unitlength\epsfbox{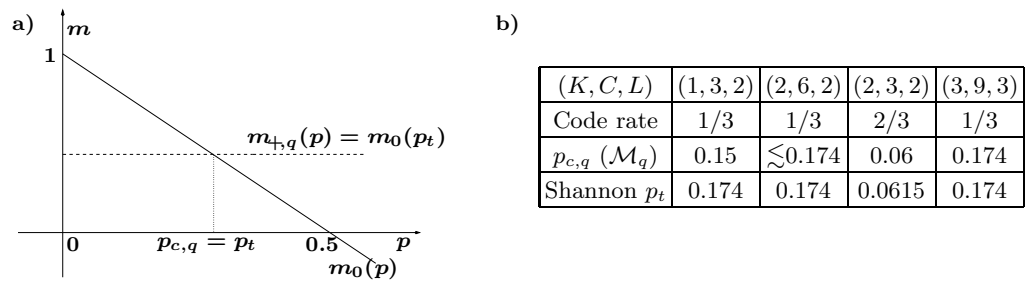}}
\put(-10, 55){\mbox{\fns \boldmath ${\rm a)}$}}
\put(- 3, 48){\mbox{\fns \boldmath $1$}}
\put(  2, 54){\mbox{\fns \boldmath $m$}}
\put(  2,  6){\mbox{\fns \boldmath $0$}}
\put( 75,  7){\mbox{\fns \boldmath $p$}}
\put( 22,  7){\mbox{\fns \boldmath $p_{c,q}=p_t$}}
\put( 60,  1){\mbox{\fns \boldmath $m_0(p)$}}
\put( 42, 30){\mbox{\fns \boldmath $m_{\e,q}(p)=m_0(p_t)$}}
\put( 55,  6){\mbox{\fns \boldmath $0.5$}}
\end{picture}
\put(-43,55){\mbox{\fns \boldmath ${\rm b)}$}}
\put(-35,30){
\begin{tabular}{|c|c|c|c|c|}
\hline
$(K,C,L)$           & $(1,3,2)$ & $(2,6,2)$ & $(2,3,2)$ & $(3,9,3)$ \\
\hline
Code rate           & $1/3    $ & $1/3    $ & $2/3    $ & $1/3    $ \\
\hline
$p_{c,q}$ ($\cM_q)$ & $0.15$&$\lesssim$0.174& $0.06   $ & $0.174  $ \\
\hline
Shannon $p_t$       & $0.174$   & $0.174  $ & $0.0615 $ & $0.174  $ \\
\hline
\end{tabular}
}
\caption{
 {\bf a)} Determining the critical noise levels $p_{c,q}$
  based on the function $\cM_q$ for MN codes with $K\geq3$ or $K=2,~L\geq3$,
  a qualitative picture.
 {\bf b)} Comparison of different critical noise level ($p_{c,q}$) estimates for
  MN codes. The numerical precision is up to the last digit for the current
  method. Shannon's limit denotes the highest theoretically achievable critical
  noise level $p_t$ for any code~\cite{shannon}. 
}
\label{fig:3}
\end{figure}
%
\section{Conclusions}
%
In this paper we  have shown how both weight and magnetization enumerators can
be calculated  using the methods of statistical physics in the case of regular
LDPC codes. We study the role played by the {\em magnetization enumerator}
$\cM(m)$ in determining the achievable critical noise level for various decoding
schemes. The formalism based on the magnetization enumerator $\cM$ offers a
intuitively simple alternative to the weight enumerator formalism used in
conjunction with typical pairs decoding in the IT literature~\cite{Aji,KNM}. The
SP based analysis employes the replica method given the very low critical values
obtained by the annealed approximation calculation. Furthermore, the powerfull
gauge theory as proposed by Nishimori~\cite{nishimori_book}, proves that the
replica symmetric assumption is correct (at least at the critical noise level),
and thus that the critical noise levels as obtained by our method are {\em
exact}.
Although we have concentrated here on the critical noise level for the BSC,
other channel types as well as other quantities of interest can be treated using
a similar formalism. The predictions for the critical noise level are more
optimistic than those reported in the IT literature, and are up to numerical
precision in agreement with those reported in~\cite{KNM}. We have also shown
that the critical noise levels for typical pairs, MAP and MPM decoding must
coincide, and we have provided an intuitive explanation to the difference
between MAP and MPM decoding. Finally, an extension of this analysis to MN codes
reveals the mechanism which allows them to saturate Shannon's limit for finite
$K\geq3$ and for $K=2,~L\geq3$ values (if impractical algorithms such as maximum
likelihood are used). This result, which is consistent with previous SP based
analyses \cite{us_prl} is considered as surprising in the IT community.

We believe that SP based analysis will provide more insight into the performance
and characteristics of random LDPC codes, complementing the analysis provided
by the methods of IT.

\vspace{3mm}
\noindent
{\footnotesize
Support by Grants-in-Aid Nos. 13680400 and 13780208 (YK),
The Royal Society and EPSRC-GR/N00562 (DS/JvM) is acknowledged.
}


\end{document}